\documentclass[11pt]{article}

\usepackage[final]{acl}

\usepackage{times}
\usepackage{latexsym}
\usepackage{amsmath}
\usepackage{enumitem}
\usepackage{booktabs}
\usepackage{multirow}
\usepackage{pgfplots}
\pgfplotsset{compat=newest}

\usepackage[T1]{fontenc}

\usepackage[utf8]{inputenc}

\usepackage{microtype}

\usepackage{inconsolata}

\usepackage{graphicx}
\usepackage[table,xcdraw]{xcolor}

\title{Pseudo2Real: Task Arithmetic for Pseudo-Label Correction in Automatic Speech Recognition}

\usepackage{xspace}
\author{Yi-Cheng Lin\thanks{These authors contributed equally.}\quad Yu-Hsuan Li Liang$^*$ \quad Hsuan Su\quad Tzu-Quan Lin \\
\textbf{Shang-Tse Chen\quad Yun-Nung Chen\quad Hung-yi Lee$^{\dagger}$}\\
National Taiwan University, Taipei, Taiwan \\
$^{\dagger}$NTU Artificial Intelligence Center of Research Excellence, Taipei, Taiwan \\
\texttt{\{f12942075, r14922013, hungyilee\}@ntu.edu.tw}}
\newcommand{\ours}{Pseudo2Real\xspace}

\begin{document}
\maketitle
\begin{abstract}
Robust ASR under domain shift is crucial because real-world systems encounter unseen accents and domains with limited labeled data. 
Although pseudo-labeling offers a practical workaround, it often introduces systematic, accent-specific errors that filtering fails to fix. 
We ask: How can we correct these recurring biases without target ground truth? 
We propose a simple parameter-space correction: in a source domain containing both real and pseudo-labeled data, two ASR models are fine-tuned from the same initialization, one on ground-truth labels and the other on pseudo-labels, and their weight difference forms a correction vector that captures pseudo-label biases.
When applied to a pseudo-labeled target model, this vector enhances recognition, achieving up to a 35\% relative Word Error Rate (WER) reduction on \textsc{AfriSpeech-200} across ten African accents with the Whisper \textsc{tiny} model. 
\end{abstract}

\begin{figure}[t]
  \centering
  \includegraphics[width=\columnwidth]{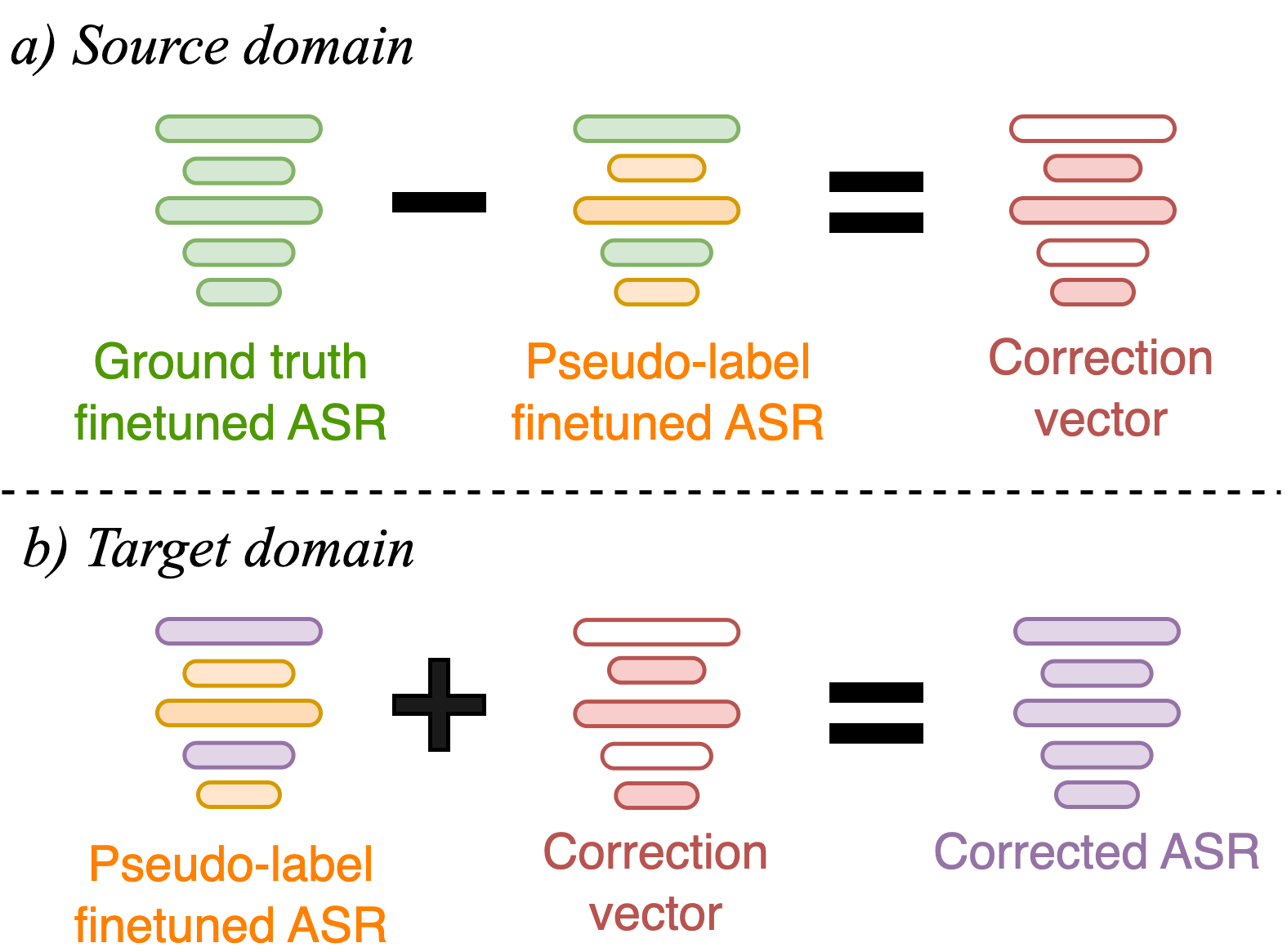}
  \caption{
  \textbf{Overview of \ours.}
  a) In the source domain, two ASR models are fine-tuned from the same pretrained initialization: 
  one using ground-truth transcripts and one using pseudo-labels. 
  Their parameter difference defines a \textcolor{red!70!black}{correction vector} that captures systematic pseudo-labeling biases. 
  b) In a new target domain, this correction vector is added to a pseudo-label fine-tuned model to produce a \textcolor{violet!70!black}{corrected ASR} that better aligns with real-label performance. Color semantics: \textcolor{green!50!black}{green} = source-domain (ground-truth) knowledge, \textcolor{orange!80!black}{orange} = pseudo-label noise, and \textcolor{violet!70!black}{purple} = target-domain knowledge.
  }
  \label{fig:overview}
\end{figure}

\section{Introduction}
ASR technologies are increasingly deployed across various domains, including smart assistants, medical transcription, and emerging low-resource accents or languages \cite{ZHANG2023106517}. 
However, when models encounter speech from new domains, labeled training data is often scarce or completely unavailable \cite{mai22_interspeech, damianos25_interspeech}. Collecting high-quality transcriptions in these settings is costly, time-consuming, and sometimes infeasible due to privacy or legal constraints \cite{bäckström2025privacyspeechtechnology, shoemate2022sotto}.

A common approach is to generate pseudo-labels using existing ASR models \cite{pmlr-v187-likhomanenko23a, 9414058}. 
However, pseudo-labels inherit the teacher model’s systematic biases, such as under-recognizing rare words, accent-driven substitutions, or domain-specific mis-segmentations \cite{9846970}.
When used for adaptation, these errors can accumulate and degrade real-world performance \cite{prakash2025betterpseudolabelingmultiasrfusion}, revealing the need for methods that automatically identify and correct structured pseudo-label errors without relying on target-domain ground truth.

In this work, we ask: \textbf{How can we mitigate systematic error patterns in ASR pseudo-labeling when no ground truth annotations are available in the target domain?} 
Prior efforts address this challenge indirectly. Teacher-student self-training improves with confidence filtering and agreement checks \cite{flynn24_interspeech, Kim_Lee_2025}, yet these strategies suppress noise without correcting the structured biases from the teacher. Iterative schedules such as Noisy Student \cite{10095704} and moving-average teacher updates \cite{zhang2024conformer1robustasrlargescale} improve pseudo-labels but require multiple passes and careful tuning, and they still propagate the teacher’s recurring mistakes.

We propose \ours, a parameter-space correction that operates without target labels, as depicted in Figure~\ref{fig:overview}. 
In source domains that provide both real transcripts and pseudo-labels, we fine-tune two models from the same backbone and form a correction vector as their difference. 
This vector captures systematic discrepancies introduced by pseudo-labeling. 
When adapting to a new target domain, the model fine-tuned on pseudo-labeled audio is adjusted by adding a scaled version of this vector, yielding a corrected ASR system that better aligns with real-label performance.
Our extensive experiments on the AfriSpeech-200 dataset demonstrate that \ours achieves consistent gains across ten African accents and multiple Whisper model sizes, including up to 35\% relative WER reduction on Whisper \textsc{tiny}. Our main contributions include:
\begin{enumerate}[nosep,leftmargin=*]
    \item We introduce \textbf{Pseudo2Real}, an effective parameter-space correction that mitigates systematic pseudo-label errors.  
    \item We extend it to \textbf{Pseudo2Real-SC}, which leverages speaker clustering to compute subgroup-specific correction vectors, thereby further enhancing robustness.  
    \item We demonstrate substantial performance improvements across accents and model scales, analyze the effect of scaling factors and the number of clusters, and provide insights into how structured pseudo-label biases can be corrected directly in parameter space.  
\end{enumerate}

\section{Related work}
\subsection{Pseudo-labeling for ASR unsupervised domain adaptation}
Work in ASR widely adopts teacher–student self-training to exploit unlabeled target audio \cite{flynn24_interspeech}. A first line of research uses a strong teacher to generate pseudo-labels, then trains a student on these labels. The quality of pseudo-labels can be improved through iterative self-training approaches such as Noisy Student \cite{park20d_interspeech, singh23b_interspeech, 10448438}, where the teacher is repeatedly updated on its own pseudo-labels with noise injection and augmentation, yielding large relative WER gains in ASR adaptation. 
A second line focuses on label quality control. Confidence filtering and agreement checks between models are used to downweight or discard unreliable segments before student training, which prevents error amplification in the target domain \cite{Kim_Lee_2025, 10.1109/TASLP.2023.3306709, pmlr-v187-likhomanenko23a}. 
A third line refines the teacher itself during adaptation \cite{10096983}. For example, KAIZEN updates the teacher as an exponential moving average of the student, yielding stronger pseudo-labels and improved unsupervised adaptation \cite{9688028}.

Our work differs in purpose and mechanism. 
Instead of only filtering or iterating on noisy pseudo-labels, we learn a reusable correction in parameter space. 
We construct a vector that captures the discrepancy between models trained on synthetic and real speech in auxiliary domains, then add this vector to a pseudo-label adapted model in a new domain to mitigate systematic error patterns without any target labels. 
This complements prior pseudo-label pipelines and can be combined with confidence filtering or iterative self-training.

\subsection{Task arithmetic in speech}


Recent works have explored task vectors (task arithmetic) \cite{ilharco2023editing} as a means to transfer capabilities between models \cite{li2025when, lin2025speechftmergingpretrainedfinetuned, huang2024multimodal, rittergutierrez25_interspeech, correlation_permutation}. In ASR, Task Vector Algebra shows that difference vectors between models trained on related settings can enable zero-shot domain adaptation and task analogy for low-resource scenarios \cite{10447848}. Extending this idea, \citet{kang24_interspeech} demonstrates that multilingual ASR can be controlled or composed across languages via simple vector addition or negation, while \citet{10890762} shows that combining task vectors from related high-resource languages improves low-resource ASR through cross-lingual transfer. LoRS-Merging \cite{zhao2025lowranksparsemodelmerging} merges language- or task-specific deltas using low-rank and sparse decomposition to enhance multilingual ASR without retraining. Building on this paradigm, SYN2REAL \cite{syn2real} defines a vector between ASR models fine-tuned on authentic versus synthetic speech and applies it to bridge the gap in acoustic signal distributions.

Our work adopts task arithmetic but targets a different problem: mitigating systematic pseudo-label errors in unsupervised ASR domain adaptation, rather than transferring general capabilities across languages, modalities, or text domains. The closest prior method is SYN2REAL \cite{syn2real}: it constructs a task vector between ASR models trained on real vs.\ synthetic audio to close the acoustic gap between TTS-generated and human speech, typically used for adapting to new text domains. In contrast, \ours constructs a correction vector between models trained on real vs.\ pseudo-labels of the same real audio to mitigate systematic label-noise biases, and applies it to new acoustic domains such as accented speech. Furthermore, while SYN2REAL relies on \emph{domain labels} to ensemble task vectors across multiple text domains, our \ours-SC variant uses automatic speaker clustering to form multiple correction vectors, requiring no domain labels in the source data.



\section{Methodology}
\subsection{Problem Formulation}
We study acoustic domain adaptation for ASR, focusing on accent as the primary axis of domain variation. Let the source domain $D_s$ consist of paired speech and text $(S_s, T_s)$, and let the target domain $D_t$ provide only unlabeled speech $S_t$. Ground-truth transcriptions $T_t$ are unavailable due to annotation cost. A common strategy is to train a teacher ASR model on $D_s$, generate pseudo-labels $\hat{T}_t$ for $S_t$, and then train a student ASR model on $(S_t, \hat{T}_t)$.

This approach is effective but suffers from \textbf{systematic error propagation}: if the teacher consistently misrecognizes rare words or accent-specific patterns, these biases are inherited by the student. Confidence filtering or re-weighting pseudo-labels can reduce noise, but cannot correct the structured error patterns that arise from teacher model biases. Our goal is therefore to automatically detect and mitigate systematic pseudo-label errors without ground-truth annotations in $D_t$.

\subsection{\ours}
\label{ssec:method_p2r}
\begin{figure}[t]
  \centering
\includegraphics[width=\columnwidth]{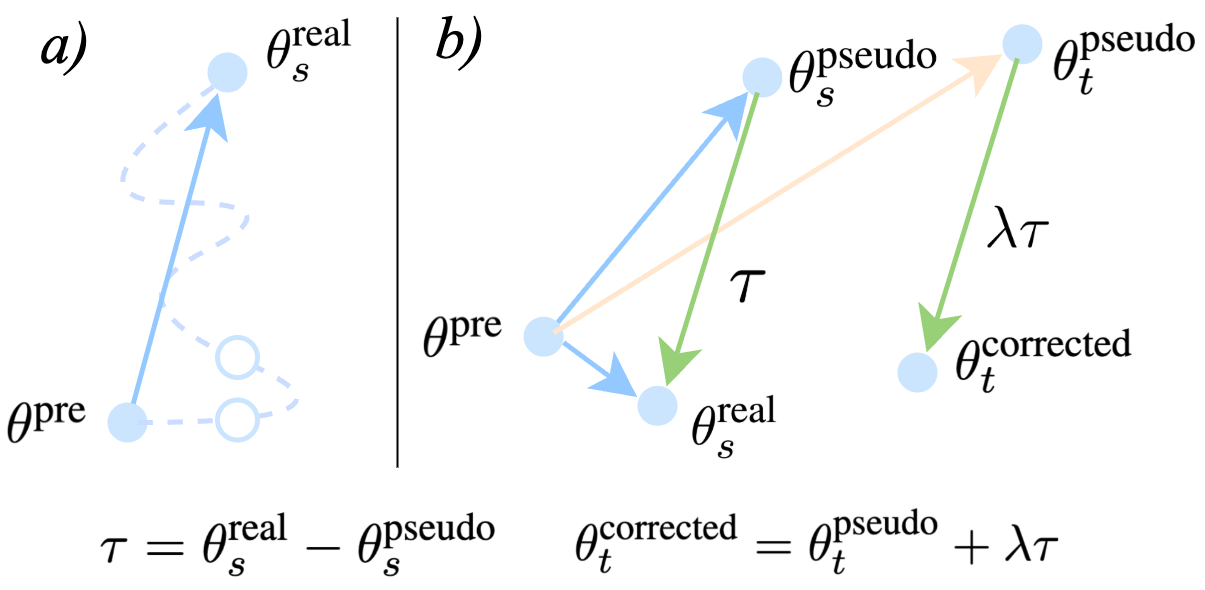}
  \caption{
\textbf{Learning and applying correction vectors in parameter space.}
  \emph{a)} A task vector is obtained by taking the difference between a pretrained model $\theta^{\text{pre}}$ and its fine-tuned version $\theta_{s}^{\text{real}}$ (or $\theta_{s}^{\text{pseudo}}$). 
  \emph{b)} In the source domain, two models are fine-tuned from the same pretrained initialization $\theta^{\text{pre}}$: one with real transcripts ($\theta_{s}^{\text{real}}$) and one with pseudo-labels ($\theta_{s}^{\text{pseudo}}$). Their difference defines the correction vector $\tau$. In a new target domain, we first obtain $\theta_{t}^{\text{pseudo}}$ by fine-tuning on pseudo-labels, then apply the correction vector to yield the final model $\theta_{t}^{\text{corrected}}$.
  }
  \label{fig:method-vector}
\end{figure}
We build on task arithmetic to design a parameter-space correction transferable across domains. A key enabling observation is \emph{linear mode connectivity} \cite{frankle2020linear, neyshabur2020transferred}: models fine-tuned from the same pre-trained initialization tend to remain within a shared low-loss region of parameter space, so that their weight differences can be interpreted as meaningful directions rather than arbitrary noise. This interpretation underlies the task-vector framework \cite{ilharco2023editing} and has been further supported by theoretical analyses showing that fine-tuning induces \emph{weight-disentangled} directions in parameter space, each governing a localized region of function space \cite{ortizjimenez2023task}. Our key observation is that, in a source domain where both real labels and pseudo-labels are available, one can learn such a direction between models trained on the two label types; this direction captures the systematic biases induced by pseudo-labels in that domain. Adding a scaled version of it to a target-domain pseudo-label-trained model shifts the parameters along the pseudo-to-real axis identified in the source. Moreover, since each such direction is defined between endpoints fine-tuned from the same initialization, task arithmetic shows that multiple directions can be linearly combined to compose their effects \cite{ilharco2023editing}, motivating the aggregation of correction vectors across speaker subgroups. These properties motivate the two variants of our method introduced below.

\paragraph{\ours: Single correction Vector.}  
Starting from the same pre-trained backbone $\theta^{\text{pre}}$, we fine-tune two student ASR models on the source domain: $\theta_{s}^{\text{real}}$, trained on $(S_s, T_s)$, and $\theta_{s}^{\text{pseudo}}$, trained on $(S_s, \hat{T}_s)$ where $\hat{T}_s$ are pseudo-labels generated by a teacher for $S_s$. The difference between these two student models defines a correction vector:
\begin{equation}
    \tau = \theta_{s}^{\text{real}} - \theta_{s}^{\text{pseudo}}.
\end{equation}
To adapt to the target domain, we fine-tune a student model on $(S_t, \hat{T}_t)$ to obtain $\theta_{t}^{\text{pseudo}}$, and then apply the correction vector:
\begin{equation}
    \theta_{t}^{\text{corrected}} = \theta_{t}^{\text{pseudo}} + \lambda \tau,
\end{equation}
where $\lambda$ is a scaling factor controlling the correction strength. This method applies a single correction vector derived from the source domain directly to the target model.

\paragraph{\ours-SC: Subgroup Correction Vectors.}  
The second variant extends this idea by recognizing that systematic pseudo-label errors may not be homogeneous across all speakers in the source domain. 
In practice, pseudo-labeling quality can vary substantially due to accent, pronunciation style, or recording conditions. 
For example, a teacher model may systematically substitute certain consonants for speakers with a specific accent, while producing relatively accurate transcriptions for speakers from another subgroup. 
If all speakers are pooled together when constructing the correction vector, these fine-grained biases may not be taken into account, weakening the correction signal.

To address this, we propose partitioning the source domain into more coherent speaker subgroups and computing subgroup-specific correction vectors. Inspired by \citet{lin25c_interspeech}, we refine the correction by exploiting speaker diversity within the source domain. We use ECAPA-TDNN\footnote{  \url{https://huggingface.co/speechbrain/spkrec-ecapa-voxceleb}} embeddings \cite{desplanques20_interspeech} to extract speaker representations for each utterance, and apply $k$-means clustering \cite{kmeans} to partition the source data into speaker subgroups. For each subgroup $c$, we fine-tune two models from $\theta^{\text{pre}}$: $\theta_{s,c}^{\text{real}}$, trained on real transcriptions of the subgroup, and $\theta_{s,c}^{\text{pseudo}}$, trained on pseudo-labels of the subgroup. We then compute a subgroup-specific correction vector:
\begin{equation}
    \tau_{c} = \theta_{s,c}^{\text{real}} - \theta_{s,c}^{\text{pseudo}}.
\end{equation}
The final correction is obtained by averaging across all $C$ clusters and applying it to the target model:
\begin{equation}
    \theta_{t}^{\text{corrected}} = \theta_{t}^{\text{pseudo}} +  \frac{\lambda}{C} \sum_{c=1}^{C} \tau_{c}.
\end{equation}
This aggregated vector captures systematic pseudo-label biases shared across speakers while preserving accent- or subgroup-specific corrections.

\subsection{Baselines}
\label{ssec:baselines}
We compare \ours against six baselines and a topline. As \emph{no-adaptation references}, \textbf{$\theta^{\text{pre}}$} applies the pretrained Whisper directly to the target domain, and \textbf{$\theta_{s}^{\text{real}}$} fine-tunes the student on source-domain ground-truth labels; together they show how far a model can go without target-domain supervision. As the \emph{pseudo-labeling baseline}, \textbf{$\theta_{t}^{\text{pseudo}}$} fine-tunes the student on target-domain pseudo-labels; this is the direct baseline that \ours aims to improve upon. Prior-art methods for pseudo-label noise fall into two groups. \emph{Label-space} methods act on the pseudo-labels: \textbf{Confidence filtering (Conf.)} retains pseudo-labels whose average word-level log-probability exceeds a development-set threshold, and \textbf{Error Correction (EC)} uses a T5 model trained on source-domain (hypothesis, ground-truth) pairs to rewrite target-domain pseudo-labels before student fine-tuning. \emph{Weight-space} methods modify the student's parameters: \textbf{Exponential Moving Average (EMA)} maintains a running average of the student's weights during training and uses it at evaluation. \ours is also a weight-space method, but uses a source-derived correction vector rather than a training-trajectory average. Finally, the \textbf{topline} fine-tunes the student on target-domain ground-truth labels as an upper-bound reference. EC and EMA implementation details are in Appendix~\ref{app:baseline_details}.




\section{Experimental Setup}

\paragraph{Dataset.} We evaluate our method on the Afro-centric benchmark \textsc{AfriSpeech-200}\footnote{ \url{https://huggingface.co/datasets/intronhealth/afrispeech-200}} \cite{afrispeech_200}, a 200-hour corpus of transcribed English speech from speakers representing 120 African accents, with explicit accent annotations. 
Accented speech remains a persistent challenge for ASR systems because strong accent variations often fall outside the distribution of large-scale pretraining corpora. 
\textsc{AfriSpeech-200} therefore provides a rigorous testbed for domain adaptation methods. 
For our experiments, we filter the corpus by accent and select the ten accents with the largest number of samples.
Not all accents include complete train, development, and test splits, so we restrict our selection to accents where all three splits are available and preserve the official split to avoid data leakage.
All utterances in these subsets are English speech; each accent label refers to the native language background of the speakers rather than the language of the transcript (e.g., ``Igbo'' denotes Igbo-accented English spoken by Igbo native speakers, not transcripts in the Igbo language).

\paragraph{Cross-Fold Validation.}
To evaluate the generalization ability of our method across diverse accents, we construct a cross-fold validation setting based on the ten most represented accents in \textsc{AfriSpeech-200}. We group these accents into two folds with comparable numbers of utterances and speakers, so that neither adaptation direction benefits from substantially more training data: \emph{fold 1} consists of \{Igbo, Swahili, Hausa, Zulu, Twi\}, and \emph{fold 2} consists of \{Yoruba, Ijaw, Afrikaans, Idoma, Setswana\}. Although all ten accents originate from African regions, they span multiple and largely unrelated language families (Niger--Congo, Afro-Asiatic, and Indo-European), and differ markedly in tonal systems, syllable structures, consonant inventories, and morphology; cross-accent adaptation across these families is therefore highly non-trivial. We list the linguistic family of each accent in Appendix~\ref{sec:language_family}. In each experiment, one fold serves as the source domain and the other as the target. The source fold provides paired speech and transcripts that are used to derive correction vectors, while the target fold provides only speech for pseudo-labeling. We then swap the roles of the folds to form the second validation round. This design ensures that the evaluation covers accents with different phonological and prosodic characteristics. 

\paragraph{Model.} We employ the Whisper family of models \cite{whisper}, which cover a wide range of capacities. Specifically, we experiment with Whisper \textsc{tiny}, \textsc{base}, \textsc{small}, \textsc{medium}, and \textsc{large-v2}. These models share the same encoder–decoder transformer architecture but differ in scale, ranging from 39M to 1.55B parameters, detailed in Table~\ref{tab:whisper_models}. All models are pre-trained on approximately 680k hours of weakly supervised speech and are widely adopted in both research and real-world applications \cite{yang2024buildingtaiwanesemandarinspoken, wu-etal-2024-codec, luo25_interspeech, 10832317}. Despite its scale and multilingual coverage, prior work has shown that its performance still degrades substantially when facing strong accent variation or domain-specific shifts \cite{10.1121/10.0024876}. Evaluating across the full model series allows us to assess the effectiveness of our correction method under both low-capacity and high-capacity regimes.

\begin{table*}[tbp]
\centering
\small
\setlength{\tabcolsep}{5pt}
\begin{tabular}{llccccccccccc}
\toprule
\textbf{Model} & \textbf{Variant} & \textbf{Igbo} & \textbf{Swahili} & \textbf{Hausa} & \textbf{Zulu} & \textbf{Twi} & \textbf{Yoruba} & \textbf{Ijaw} & \textbf{Afrikaans} & \textbf{Idoma} & \textbf{Setswana} & \textbf{Avg.} \\
\midrule
\multirow[c]{8}{*}{Tiny}  & $\theta^{\text{pre}}$ & 93.2 & 77.7 & 142.8 & 79.0 & 73.3 & 94.8 & 191.2 & 54.4 & 182.9 & 75.6 & 106.5 \\
    & $\theta_{s}^{\text{real}}$ & \textbf{60.1} & 67.3 & 135.3 & 70.8 & 53.8 & 105.0 & 155.8 & 51.8 & 128.2 & 54.4 & 88.2 \\
    & $\theta_{t}^{\text{pseudo}}$ & 61.8 & 70.5 & 119.6 & 75.2 & 59.4 & 112.3 & 157.7 & 52.5 & 129.0 & 55.0 & 89.3 \\
    & conf. & 73.0 & 58.7 & 147.7 & \textbf{56.0} & 57.7 & 97.8 & 194.6 & 52.0 & 107.2 & \textbf{41.9} & 88.7 \\
    & EC   & 81.7 & 64.5 & 93.0 & 79.4 & 75.6 & 122.2 & 137.2 & 47.8 & 83.7 & 53.1 & 83.8 \\
    & EMA  & 88.2 & 62.4 & 131.8 & 79.4 & 74.7 & 122.0 & 167.1 & 57.4 & 140.1 & 63.8 & 98.7 \\
    & {\color{red} ours} & 60.3 & \textbf{51.9} & \textbf{78.8} & 67.0 & \textbf{45.5} & \textbf{61.3} & \textbf{61.3} & \textbf{45.4} & \textbf{60.7} & 44.8 & \textbf{57.7} \\
    & \textcolor{gray}{topline} & \textcolor{gray}{65.7} & \textcolor{gray}{45.8} & \textcolor{gray}{54.1} & \textcolor{gray}{59.7} & \textcolor{gray}{88.0} & \textcolor{gray}{56.1} & \textcolor{gray}{58.1} & \textcolor{gray}{41.4} & \textcolor{gray}{112.4} & \textcolor{gray}{44.8} & \textcolor{gray}{62.6} \\
\midrule
\multirow[c]{8}{*}{Small} & $\theta^{\text{pre}}$ & 56.4 & 49.6 & 60.6 & 49.7 & 49.7 & 62.8 & 63.2 & 41.3 & 66.0 & 49.7 & 54.9 \\
    & $\theta_{s}^{\text{real}}$ & 52.9 & \textbf{39.7} & 73.8 & 40.7 & 37.5 & 56.9 & 52.3 & 36.0 & 55.6 & 40.3 & 48.6 \\
    & $\theta_{t}^{\text{pseudo}}$ & 53.1 & 39.9 & \textbf{58.2} & 41.1 & 37.5 & 57.2 & 52.4 & 36.4 & 55.8 & 40.7 & 47.2 \\
    & conf. & 52.1 & 43.0 & 78.3 & 42.5 & 38.2 & 50.2 & 49.9 & 36.2 & 55.8 & 41.9 & 48.8 \\
    & EC   & 50.6 & 40.9 & 63.1 & 44.3 & 43.8 & 50.8 & 137.0 & \textbf{30.6} & 55.5 & 39.8 & 55.6 \\
    & EMA  & 62.9 & 47.3 & 66.8 & 49.4 & 52.9 & 62.0 & 61.8 & 41.0 & 63.7 & 48.6 & 55.6 \\
    & {\color{red} ours} & \textbf{46.5} & 42.3 & 78.0 & \textbf{38.5} & \textbf{35.0} & \textbf{44.1} & \textbf{47.1} & 32.4 & \textbf{48.9} & \textbf{37.1} & \textbf{45.0} \\
    & \textcolor{gray}{topline} & \textcolor{gray}{49.6} & \textcolor{gray}{35.7} & \textcolor{gray}{52.5} & \textcolor{gray}{36.8} & \textcolor{gray}{35.0} & \textcolor{gray}{41.3} & \textcolor{gray}{45.4} & \textcolor{gray}{29.5} & \textcolor{gray}{47.7} & \textcolor{gray}{36.2} & \textcolor{gray}{41.0} \\
\bottomrule
\end{tabular}
\caption{\textbf{WER (\%) on ten accented English target domains.} Results are shown for Whisper \textsc{tiny} and Whisper \textsc{small} under eight adaptation settings. Lower is better. The best performance is \textbf{bolded}.}
\label{tab:results_same_ts}
\end{table*}

\paragraph{Training.} In our experiments, we fully fine-tune the Whisper \textsc{small} and \textsc{tiny} models as student models using the AdamW optimizer \cite{adamw} with a learning rate of $3\times10^{-5}$ and a weight decay of 0.1.
Training is conducted for up to 40K update steps with a linear warmup of 500 steps.
Each model is trained with a batch size of 16, using mixed-precision (FP16) to reduce memory consumption. Evaluation is performed every 50 steps using the word error rate (WER) metric with greedy decoding.
For \emph{task arithmetic}, we use the entire model parameter to compute the correction vector. 
The scaling factor $\lambda$ is selected using the source-domain development sets: we perform a simple grid search over $\lambda \in \{0.1, 0.2, 0.3, ..., 1.0\}$ and choose the value that minimizes WER on the held-out source development set. 
All experiments are conducted on a single NVIDIA V100 GPU, totaling approximately 500 GPU-hours.

\begin{table*}[t]
\centering
\small
\begin{tabular}{l l l c c c c c c}
\toprule
\textbf{Student} & \textbf{Teacher} & \textbf{Variant} & \textbf{Igbo} & \textbf{Swahili} & \textbf{Hausa} & \textbf{Zulu} & \textbf{Twi} & \textbf{Avg.} \\
\midrule
\multirow[c]{12}{*}{Tiny} 
 & -- & $\theta^{\text{pre}}$       & 93.2 & 77.7 & 142.8 & 79.0 & 73.3 & 93.20 \\ 
\cmidrule(lr){2-9}
 & \multirow{3}{*}{Base}& $\theta_{t}^{\text{pseudo}}$ & 73.9 & 55.6 & 94.9 & 82.8 & 64.6 & 74.36 \\
 &  & Pseudo2Real   & 52.5 & 50.6 & 76.6 & 66.1 & 44.4 & 58.04 \\
 &  & Improvement (\%)   & \textbf{\textcolor{green!50!black}{+29.0}} & \textbf{\textcolor{green!50!black}{+9.0}} & \textbf{\textcolor{green!50!black}{+19.2}} & \textbf{\textcolor{green!50!black}{+20.2}} & \textbf{\textcolor{green!50!black}{+31.3}} & \textbf{\textcolor{green!50!black}{+21.7}} \\
\cmidrule(lr){2-9}
  & \multirow{3}{*}{Small}& $\theta_{t}^{\text{pseudo}}$ & 59.88 & 50.44 & 66.59 & 65.58 & 49.66 & 58.43 \\
 &  & Pseudo2Real & 52.58 & 49.15 & 73.25 & 48.33 & 44.24 & 53.51 \\
 &  & Improvement (\%)   & \textbf{\textcolor{green!50!black}{+12.2}} & \textbf{\textcolor{green!50!black}{+2.6}} & \textbf{\textcolor{red}{-10.0}} & \textbf{\textcolor{green!50!black}{+26.3}} & \textbf{\textcolor{green!50!black}{+10.9}} & \textbf{\textcolor{green!50!black}{+8.4}} \\
\cmidrule(lr){2-9}
  & \multirow{3}{*}{Medium}& $\theta_{t}^{\text{pseudo}}$ & 58.44 & 51.48 & 72.54 & 49.81 & 46.39 & 55.73 \\
 &  & Pseudo2Real  & 65.19 & 46.78 & 64.66 & 47.42 & 42.33 & 53.28 \\
 &  & Improvement (\%)   & \textbf{\textcolor{red}{-11.5}} & \textbf{\textcolor{green!50!black}{+9.1}} & \textbf{\textcolor{green!50!black}{+10.9}} & \textbf{\textcolor{green!50!black}{+4.8}} & \textbf{\textcolor{green!50!black}{+8.8}} & \textbf{\textcolor{green!50!black}{+4.4}} \\
\cmidrule(lr){2-9}
  & \multirow{3}{*}{Large}& $\theta_{t}^{\text{pseudo}}$ & 68.19 & 54.45 & 103.39 & 50.09 & 50.00 & 65.22 \\
 &  & Pseudo2Real  & 51.77 & 43.26 & 56.06 & 48.44 & 42.89 & 48.48 \\
 &  & Improvement (\%)   & \textbf{\textcolor{green!50!black}{+24.1}} & \textbf{\textcolor{green!50!black}{+20.5}} & \textbf{\textcolor{green!50!black}{+45.8}} & \textbf{\textcolor{green!50!black}{+3.3}} & \textbf{\textcolor{green!50!black}{+14.2}} & \textbf{\textcolor{green!50!black}{+21.6}} \\
\midrule
\multirow[c]{9}{*}{Small} 
  & -- & $\theta^{\text{pre}}$  & 56.4 & 49.6 & 60.6 & 49.7 & 49.7 & 53.20 \\
\cmidrule(lr){2-9}
 & \multirow{3}{*}{Medium}& $\theta_{t}^{\text{pseudo}}$ & 46.79 & 39.35 & 50.86 & 40.01 & 36.68 & 42.74 \\
 &  & Pseudo2Real  & 40.31 & 37.25 & 50.86 & 39.27 & 34.76 & 40.49 \\
 &  & Improvement (\%)   & \textbf{\textcolor{green!50!black}{+13.8}} & \textbf{\textcolor{green!50!black}{+5.3}} & \textbf{+0.0} & \textbf{\textcolor{green!50!black}{+1.8}} & \textbf{\textcolor{green!50!black}{+5.2}} & \textbf{\textcolor{green!50!black}{+5.2}} \\
\cmidrule(lr){2-9}
  & \multirow{3}{*}{Large}& $\theta_{t}^{\text{pseudo}}$ & 46.68 & 43.85 & 55.02 & 40.96 & 38.37 & 44.98 \\
 &  & Pseudo2Real & 40.14 & 45.32 & 65.33 & 39.76 & 36.23 & 45.36 \\
 &  & Improvement (\%)   & \textbf{\textcolor{green!50!black}{+14.0}} & \textbf{\textcolor{red}{-3.4}} & \textbf{\textcolor{red}{-18.8}} & \textbf{\textcolor{green!50!black}{+2.9}} & \textbf{\textcolor{green!50!black}{+5.6}} & \textbf{\textcolor{green!50!black}{+0.1}} \\
\bottomrule
\end{tabular}
\caption{\textbf{WER (\%) on five accented English target domains from \textsc{AfriSpeech-200}, under different teacher model sizes.} Each \textit{Improvement} row reports the relative improvement (\%) of our Pseudo2Real method over pseudo-labeled fine-tuning. \textbf{\textcolor{green!50!black}{Green}} indicates gains (lower WER); \textbf{\textcolor{red}{red}} indicates degradation.}
\label{tab:teacher_size}
\end{table*}

\section{Result}
\subsection{Can \ours improve ASR performance?}
We begin by examining whether the application of \ours can improve ASR performance in the cross-validation setting, compared to the baselines introduced in \S\ref{ssec:baselines}.

In our main experiments, we evaluate the case where the teacher model and the student model are identical. The results in Table~\ref{tab:results_same_ts} show that both baseline methods underperform significantly across all accent domains. The pretrained models $\theta^{\text{pre}}$ perform poorly, especially on low-resource accents such as Hausa, Ijaw, and Idoma, with average WERs of 106.5 and 54.9 for Whisper \textsc{tiny} and \textsc{small}, respectively. Fine-tuning on pseudo-labeled data $\theta_{t}^{\text{pseudo}}$ provides noticeable gains, reducing the average WER by 17.2 points for \textsc{tiny} and 7.7 points for \textsc{small}, but large error rates persist due to systematic biases in the pseudo-labels.

In contrast, \textbf{Pseudo2Real} achieves substantial improvements across all accents, demonstrating the effectiveness of parameter-space correction. For Whisper \textsc{tiny}, Pseudo2Real lowers the average WER from 89.3 to 57.7, representing a 35\% relative improvement over pseudo-label fine-tuning. For Whisper \textsc{small}, the average WER decreases from 47.2 to 45.0, bringing performance much closer to the topline trained on labeled target data. 

Pseudo2Real shows particularly strong gains on challenging accents such as \textbf{Ijaw}, \textbf{Idoma}, and \textbf{Yoruba}, where WER is reduced by up to 50 points compared with pseudo-label fine-tuning. These improvements indicate that the correction vector effectively captures accent-specific pronunciation patterns and mitigates systematic pseudo-labeling errors that standard fine-tuning fails to address.

Interestingly, Pseudo2Real occasionally outperforms the topline trained on labeled target data. 
For instance, Whisper \textsc{tiny} outperforms the topline on \textbf{Igbo}, \textbf{Twi}, and \textbf{Idoma}, while Whisper \textsc{small} does so on \textbf{Igbo}. This behavior suggests that parameter-space correction not only compensates for pseudo-label noise but also transfers beneficial regularities learned from other domains, such as better acoustic normalization and pronunciation consistency. In these cases, the correction vector serves as a form of cross-domain regularization, enabling the adapted model to generalize more effectively than models trained solely on limited labeled target data.

\paragraph{Why does \ours sometimes beat the topline?}
We compare error types on the two accents where \ours beats the topline in the tiny$\rightarrow$tiny setting, \textbf{Twi} and \textbf{Igbo} (Table~\ref{tab:err_analysis}). On both accents, \ours makes fewer \emph{insertion} errors than the topline (most strikingly, 262 vs.\ 670 on Twi), while substitutions and deletions are comparable. Since insertions are common hallucination patterns that small target-domain fine-tuning rarely fully suppresses, this reduction suggests that the correction vector transfers additional anti-hallucination regularization from the source domain.

\begin{table}[h]
\centering
\small
\setlength{\tabcolsep}{5pt}
\begin{tabular}{llccc}
\toprule
\textbf{Accent} & \textbf{Model} & \textbf{Sub.} & \textbf{Ins.} & \textbf{Del.} \\
\midrule
\multirow{2}{*}{Twi}  & \ours    & 238   & \textbf{262}  & 11  \\
                      & Topline  & 251   & 670           & 4   \\
\midrule
\multirow{2}{*}{Igbo} & \ours    & 1933  & \textbf{3577} & 205 \\
                      & Topline  & 2004  & 3748          & 189 \\
\bottomrule
\end{tabular}
\caption{Error breakdown (substitutions, insertions, deletions) for the tiny$\rightarrow$tiny setting on two accents where \ours surpasses the topline. \ours consistently reduces insertion errors, while substitution and deletion counts remain comparable.}
\label{tab:err_analysis}
\end{table}

\subsection{How does \ours perform across different teacher model sizes?}

An important question is whether \ours can generalize across different ASR model sizes rather than be tied to a specific backbone. To examine this, we consider settings where the student and teacher may differ in capacity. In Table~\ref{tab:teacher_size}, we report results on five accents in fold 1.  

For the Whisper \textsc{tiny} student, the strongest average gains occur with \textsc{base} and \textsc{large} teachers, yielding +21.7\% and +21.6\% relative improvements, respectively. The \textsc{base} teacher produces consistent reductions across accents, including \textbf{Igbo} (+29.0\%) and \textbf{Twi} (+31.3\%), while the \textsc{large} teacher achieves the largest single-accent gain on \textbf{Hausa} (+45.8\%). Using \textsc{small} or \textsc{medium} teachers leads to smaller average gains (+8.4\% and +4.4\%), with mixed outcomes such as a degradation on \textbf{Igbo} for and \textsc{medium} teacher (-11.5\%) but a strong improvement on \textbf{Zulu} for \textsc{small} teacher (+26.3\%).

For the Whisper \textsc{small} student, improvements are more modest overall. Pairing with a \textsc{medium} teacher yields a +5.2\% average relative improvement, with steady gains on nearly all accents. The \textsc{large} teacher yields nearly identical performance to the baseline on average (around +0.1\%), showing mixed results across accents. It improves on \textbf{Igbo} (+14.0\%) and \textbf{Setswana} (+5.6\%) but degrades notably on \textbf{Hausa} (-18.8\%).

Overall, \ours is effective across a range of teacher sizes, but the magnitude of improvement depends on the teacher–student pairing and the target accent.  

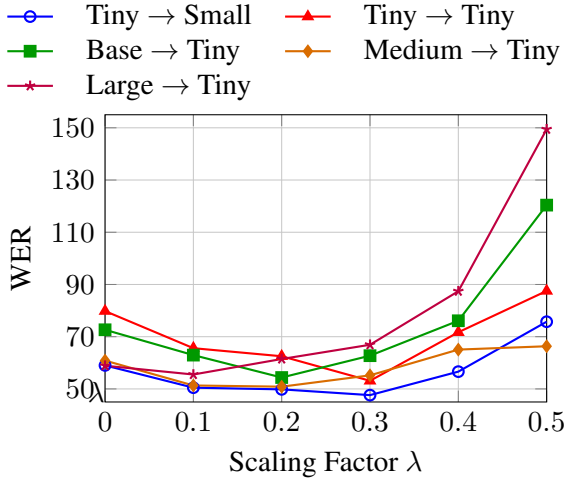
\begin{figure}[t]
  \centering
  \begin{tikzpicture}
    \begin{axis}[
      width=0.96\linewidth,
      height=0.7\linewidth,
      xlabel={Scaling Factor $\lambda$},
      ylabel={WER},
      xmin=0, xmax=0.5,
      ymin=45, ymax=155,
      xtick={0,0.1,0.2,0.3,0.4,0.5},
      ytick={50,70,90,110,130,150},
      tick label style={/pgf/number format/fixed},
      grid=both,
      grid style={line width=.2pt, draw=gray!30},
      major grid style={line width=.3pt, draw=gray!45},
      legend style={
        at={(0.41,1.02)}, anchor=south, legend columns=2,
        draw=none, fill=none, column sep=10pt
      },
      legend cell align=left
    ]

      \addplot+[mark=o, thick, blue, mark options={draw=blue, fill=blue}]
        coordinates {
          (0.0,59.002) (0.1,50.474) (0.2,49.83) (0.3,47.628) (0.4,56.62) (0.5,75.734)
        };
      \addlegendentry{Tiny $\rightarrow$ Small}

      \addplot+[mark=triangle*, thick, red, mark options={draw=red, fill=red}]
        coordinates {
          (0.0,79.814) (0.1,65.602) (0.2,62.512) (0.3,53.11) (0.4,71.7) (0.5,87.522)
        };
      \addlegendentry{Tiny $\rightarrow$ Tiny}

      \addplot+[mark=square*, thick, green!60!black, mark options={draw=green!60!black, fill=green!60!black}]
        coordinates {
          (0.0,72.65) (0.1,62.948) (0.2,54.366) (0.3,62.728) (0.4,76.1) (0.5,120.39)
        };
      \addlegendentry{Base $\rightarrow$ Tiny}

      \addplot+[mark=diamond*, thick, orange!85!black]
        coordinates {
          (0.0,60.856) (0.1,51.318) (0.2,50.86) (0.3,55.2) (0.4,65.062) (0.5,66.34)
        };
      \addlegendentry{Medium $\rightarrow$ Tiny}

      \addplot+[mark=star, thick, purple]
        coordinates {
          (0.0,58.886) (0.1,55.52) (0.2,61.458) (0.3,66.91) (0.4,87.41) (0.5,149.422)
        };
      \addlegendentry{Large $\rightarrow$ Tiny}

        {WER vs. Scaling Factor ($\lambda$)};
    \end{axis}
  \end{tikzpicture}
  \caption{\textbf{WER vs. scaling factor ($\lambda$).} Each curve corresponds to a different teacher–student pairing. Here, the arrow ($\rightarrow$) denotes that pseudo-labels are generated by the teacher ASR model on the left and used to fine-tune the student model on the right (e.g., \textsc{large}$\rightarrow$\textsc{tiny} means pseudo-labels are produced by the \textsc{large} teacher, and the \textsc{tiny} student's parameters are then adjusted using the \ours correction vector). WER values are averaged over the five fold-1 accents (Igbo, Swahili, Hausa, Zulu, Twi).
Lower WER indicates better performance.}
  \label{fig:wer-scaling-tiny}
\end{figure}

\begin{table*}[t]
\centering
\small
\begin{tabular}{l l c c c c c c}
\toprule
\textbf{Teacher} & \textbf{Variant} & \textbf{Igbo} & \textbf{Swahili} & \textbf{Hausa} & \textbf{Zulu} & \textbf{Twi} & \textbf{Avg.} \\
\midrule
\multirow[c]{3}{*}{Medium} & Pseudo2Real & 40.3 & 37.3 & 63.4 & 39.3 & 35.2 & 43.10 \\
       & Pseudo2Real-SC          & 41.1 & 37.3 & 49.6 & 38.9 & 35.4 & 40.46 \\
       & Improvement (\%) & \textbf{\textcolor{red}{-2.0}} & \textbf{0.0} & \textbf{\textcolor{green!50!black}{+21.8}} & \textbf{\textcolor{green!50!black}{+1.0}} & \textbf{\textcolor{red}{-0.6}} & \textbf{\textcolor{green!50!black}{+4.0}} \\
\midrule
\multirow[c]{3}{*}{Large}  & Pseudo2Real & 40.1 & 45.3 & 65.3 & 39.8 & 36.2 & 45.34 \\
       & Pseudo2Real-SC          & 40.3 & 40.8 & 53.9 & 39.2 & 35.4 & 41.92 \\
       & Improvement (\%) & \textbf{\textcolor{red}{-0.5}} & \textbf{\textcolor{green!50!black}{+9.9}} & \textbf{\textcolor{green!50!black}{+17.5}} & \textbf{\textcolor{green!50!black}{+1.5}} & \textbf{\textcolor{green!50!black}{+2.2}} & \textbf{\textcolor{green!50!black}{+6.1}} \\
\bottomrule
\end{tabular}
\caption{\textbf{Comparison between Pseudo2Real and its subgroup clustering variant (SC).} The \textbf{Improvement} row reports the relative change (\%) of +SC over Pseudo2Real. \textbf{\textcolor{green!50!black}{Green}} indicates gains (lower WER), \textbf{\textcolor{red}{red}} indicates degradation. Lower is better.}
\label{tab:ensemble}
\end{table*}

\subsection{What is the impact of the scaling factor $\lambda$?}

We next examine how the scaling factor $\lambda$ affects the magnitude of the applied correction vector. As described in \S~\ref{ssec:method_p2r}, $\lambda$ controls how strongly the Pseudo2Real vector influences the target model parameters. Figure~\ref{fig:wer-scaling-tiny} presents the relationship between WER and $\lambda$ for five transfer settings involving different teacher--student combinations, with $\lambda$ ranging from 0.0 (no correction) to 0.5 (strong correction). The reported WER values are averaged over the same five accents used in Table~\ref{tab:teacher_size}.    

Across all transfer settings, we observe a U-shaped trend. As $\lambda$ increases from 0, WER first decreases, reaching its minimum around $\lambda=0.2$–$0.3$, and then rises sharply for larger $\lambda$ values. This pattern suggests that small scaling factors yields the best balance between correction strength and stability, whereas excessively large $\lambda$ values can lead to over-correction and degraded accuracy. For instance, the \textsc{tiny}$\rightarrow$\textsc{tiny} and \textsc{tiny}$\rightarrow$\textsc{small} perform best at $\lambda=0.3$, while excessive scaling beyond this point causes performance degradation. Similar behavior is observed in the \textsc{base}$\rightarrow$\textsc{tiny} and \textsc{medium}$\rightarrow$\textsc{tiny} settings. However, the \textsc{large}$\rightarrow$\textsc{tiny} case shows greater instability at high $\lambda$, likely because the strong correction signal from a large teacher is difficult for a small student to absorb.   

Importantly, even small scaling factors ($\lambda=0.1$–$0.2$) consistently improve performance compared to the uncorrected case ($\lambda=0.0$). This demonstrates that applying a mild parameter-space correction is generally beneficial and robust across different model configurations. Overall, the results confirm that the scaling factor $\lambda$ plays a critical role in balancing correction strength and model stability.

\section{\ours-SC}
\subsection{How is the \ours-SC compared with the simple correction vector?}

We now investigate whether ensembling multiple correction vectors yields further improvements over a single correction vector. We evaluate two Whisper teacher sizes (\textsc{large}, \textsc{medium}) paired with the Whisper \textsc{small} student, comparing \ours with its ensemble variant in Table~\ref{tab:ensemble}. We focus on these pairings because the student benefits only modestly from a single correction vector when paired with high-capacity teachers (+5.2\% for \textsc{medium} and +0.1\% for \textsc{large} on average; Table~\ref{tab:teacher_size}), with pronounced per-accent variation. This suggests that pseudo-label biases in these settings are more heterogeneous across speakers, making them the most informative cases for testing whether subgroup clustering adds further benefit. We use 8 clusters for k-means.

The results show that \ours-SC generally maintains or improves performance relative to the single correction vector, with the magnitude of gains depending on the teacher size and target accent. For the \textsc{medium} teacher, \ours-SC achieves an average 4.0\% relative improvement, driven primarily by large gains on \textbf{Hausa} (+21.8\%), while other accents remain stable. With the \textsc{large} teacher, the ensemble variant produces a stronger average gain of 6.1\%, including notable improvements on \textbf{Hausa} (+17.5\%) and \textbf{Swahili} (+9.9\%). These results indicate that averaging subgroup-specific correction vectors can enhance robustness by capturing complementary correction patterns from diverse speaker groups.

However, improvements are not universal. \textbf{Igbo} shows slight degradation with both the \textsc{large} and \textsc{medium} teachers, indicating that excessive averaging can weaken accent-specific corrections. Overall, \ours-SC generally improves on the single correction vector, especially when the teacher has sufficient capacity to model heterogeneous speaker variations.

\begin{figure}[t]
  \centering
  \begin{tikzpicture}
    \begin{axis}[
      width=\linewidth,
      height=0.55\linewidth,
      xlabel={Number of K-means Clusters},
      ylabel={WER},
      xmin=0.5, xmax=8.5,
      ymin=40, ymax=47,
      xtick={1,2,4,8},
      ytick={41,42,43,44,45,46,47},
      grid=both,
      grid style={line width=.2pt, draw=gray!30},
      major grid style={line width=.3pt, draw=gray!45},
      tick label style={/pgf/number format/fixed},
      legend style={
        at={(0.5,1.02)}, anchor=south, legend columns=1,
        draw=none, fill=none
      },
      legend cell align=left
    ]

      \addplot+[
        mark=o,
        thick,
        color=blue,
        mark options={draw=blue, fill=blue}
      ]
        coordinates {
          (1,45.356)
          (2,44.336)
          (4,43.676)
          (8,41.944)
        };

    \end{axis}
  \end{tikzpicture}

  \caption{\textbf{WER vs. number of K-means clusters} for the \textsc{large}$\rightarrow$\textsc{small} setting. WER values are averaged over the five fold-1 accents (Igbo, Swahili, Hausa, Zulu, Twi).
  Increasing the number of clusters improves adaptation quality (lower WER).}
  \label{fig:kmeans-large-small}
\end{figure}
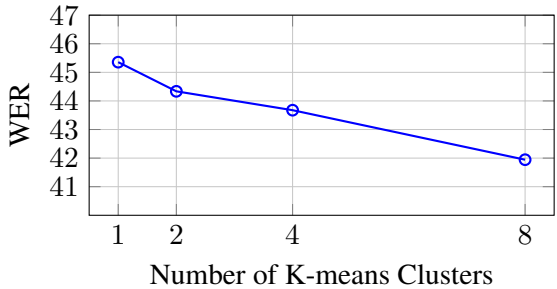

\begin{table*}[t]
\centering
\small
\setlength{\tabcolsep}{6pt}
\renewcommand{\arraystretch}{1.25}
\begin{tabular}{p{15.4cm}}
\toprule
\textbf{Example and Description} \\
\midrule
\textbf{Training set Ground Truth (Ijaw):} The codes used for the four-needle telegraph are not known, and none of the equipment has \textcolor{green!50!black}{survived}. \\
\textbf{Teacher Pseudo-label:} because used for the 4FN2 telegraph, I'm not known command, and I'm not the equipment \textcolor{red}{as a vif} \\
\textbf{Error Type:} Acoustic confusion—the teacher misinterprets the phonetic pattern of the word \textcolor{green!50!black}{``survived''} as the acoustically similar but meaningless phrase \textcolor{red}{``as a vif''}. \\[3pt]
\midrule
\textbf{Testing set Ground Truth:} If the child \textcolor{green!50!black}{survives}, he or she should be monitored for the later appearance of colonic polyps.\\
\textbf{Pretrained Tiny:} if the child \textcolor{green!50!black}{survives} he or she should be monitored for the data appearance of colonical \\
\textbf{Tiny (Pseudolabel):} if the child \textcolor{red}{is a vice} he or she should be monitored for the later appearance of colonic politics \\
\textbf{Tiny (Pseudo2Real):} if the child \textcolor{green!50!black}{survives} he or she should be monitored for the later appearance of colonic politics \\
\textbf{Error Mitigation:} Pseudo2Real restores the correct lexical meaning “survives,” correcting the acoustic corruption inherited from the teacher.\\
\bottomrule
\end{tabular}
\caption{\textbf{Qualitative examples} showing how \ours corrects systematic pseudo-label errors.  
Teacher errors (top) propagate to the student trained on pseudo-labels, while Pseudo2Real effectively suppresses these patterns and restores the intended meaning.  
\textcolor{red}{Red} = error; \textcolor{green!50!black}{green} = corrected token.}
\label{tab:case_study}
\end{table*}

\subsection{Ablation on the number of clusters}

We further analyze how the number of $k$-means clusters used in \ours-SC affects adaptation performance.  Figure~\ref{fig:kmeans-large-small} presents the results for the \textsc{large}$\rightarrow$\textsc{small} transfer setting, where the number of clusters $k$ varies from 1 (no clustering) to 8. The WER values are averaged across the same five accents used in previous experiments.

As the number of clusters increases, WER decreases. Using a single cluster corresponds to the standard Pseudo2Real setting, yielding a WER of 45.36. As $k$ increases to 2 and 4, the WER gradually decreases to 44.34 and 43.68, respectively, and reaches the lowest value of 41.94 at $k=8$. This trend suggests that finer clustering enables the model to capture more localized speaker- or accent-specific correction patterns, resulting in improved generalization to target-domain speech.

However, increasing the number of clusters raises computational cost, since two ASR models (real and pseudo) must be trained per cluster. In practice, moderate values such as $k=4$–$8$ provide a good balance between performance and efficiency. Overall, the ablation shows that leveraging speaker diversity via clustering enhances the effectiveness of parameter-space correction in \ours.

\section{Case study}
To better understand how Pseudo2Real corrects systematic pseudo-labeling errors, we present a qualitative example in Table~\ref{tab:case_study} that compares transcriptions from the teacher, student, and corrected models against the ground truth. More case studies can be found in Table~\ref{tab:additional_case_study}.

The teacher produces nonsensical output (“as a vif”), and the fine-tuned student inherits part of this lexical confusion (“is a vice”). Pseudo2Real replaces the incorrect token with the correct verb “survives,” recovering the intended meaning while keeping the rest of the sentence intact. This indicates that the correction vector adjusts the model’s internal representations to reduce systematic substitution errors commonly found in pseudo-labeling.  




\section{Conclusion}
This work proposed \textbf{\ours}, a parameter-space correction method that mitigates systematic pseudo-label errors in ASR domain adaptation without requiring target-domain labels. Experiments on \textsc{AfriSpeech-200} across ten African accents and multiple Whisper sizes show consistent gains, achieving up to 35\% relative WER reduction on Whisper \textsc{tiny} and occasionally surpassing topline models trained with true labels. We also introduced \textbf{\ours-SC}, which yields additional improvements in several teacher–student settings. Future work includes extending \ours to multilingual and spontaneous speech settings, exploring the dynamic scaling of correction strength, analyzing the interpretability of learned correction vectors, developing cluster-conditional correction vectors that can be selected or mixed adaptively at inference time based on speaker characteristics, and scaling \ours-SC to larger numbers of clusters ($k>8$) to study whether finer-grained subgroup partitioning yields further gains.
\section{Limitation}

\paragraph{Source domain supervision} Our approach assumes access to at least one source domain with paired speech and ground-truth transcriptions in order to construct the correction vector(s). 
If the available source supervision is too small, unrepresentative, or collected under markedly different conditions, the estimated vector may underfit or encode mismatched biases, which can limit transfer to the target accents.

\paragraph{Pseudo-label assumption}
The method relies on an implicit stationarity assumption: systematic biases that appear in pseudo-labels for the source domain are assumed to recur in the target domain.
When teacher errors are highly accent-specific or driven by channel and recording conditions that do not overlap with the source, the correction may be weak or even counterproductive.
Relatedly, we observed that the scaling factor $\lambda$ must be tuned, and excessive scaling can degrade WER.
Although we tune $\lambda$ only on held-out source development data, this still introduces a hyperparameter that may not transfer perfectly to new deployments.

\paragraph{Source composition.}
Our experiments rely on a fixed two-fold split, so robustness to alternative source-domain compositions (e.g., more folds, different mixes of accents or language families) is not directly characterized. This is partly bounded by data availability: most public ASR corpora (e.g., Common Voice) lack consistent or reliable accent annotations, making controlled cross-accent experiments difficult. \textsc{AfriSpeech-200} is one of the few datasets that provides curated accent labels with complete train/dev/test splits, which guided our selection. Investigating the sensitivity of the correction vector to source-fold composition, and extending the protocol to more source partitions, is an important direction for future work.

\paragraph{Language} Our experiments focus on English accents within \textsc{AfriSpeech-200}.
Generalization to other languages and domains beyond reading or conversational speech is not validated here and remains future work.

\paragraph{Accent Representation} Our experiments focus on English accents within \textsc{AfriSpeech-200}. 
We filtered to accents that provide complete train, development, and test splits to ensure a fair protocol, but this choice may bias the evaluation toward better-represented accents and does not cover underrepresented varieties or code-switching scenarios. 
Generalization to other languages, domains beyond read or conversational speech, or far-field conditions is not validated here and remains future work.

\section{Ethical considerations}
This work focuses on improving the robustness of ASR through parameter-space correction of pseudo-labeling errors. 
The research is primarily methodological and does not involve the collection of new speech data or the deployment of real-world systems. 
Nevertheless, several potential risks and ethical considerations merit discussion.

\paragraph{Bias and fairness.} 
ASR systems often exhibit disparities in accuracy across accents, dialects, and demographic groups \cite{10887767, 10.1007/978-3-031-21707-4_30, 10314895}. 
While our method aims to mitigate such disparities by improving adaptation to underrepresented accents, it may also amplify biases present in the teacher models or source-domain data. 
We encourage practitioners to evaluate model fairness carefully across linguistic and demographic subgroups when applying this technique, and to accompany adaptation with representative validation datasets.

\paragraph{Privacy and data use.}
Our experiments rely on publicly released corpora with consented speech recordings. 
No personally identifiable information or private data is used. 
However, adaptation methods in general could be misapplied to voice data collected without consent. 
Researchers and practitioners should ensure compliance with data protection regulations and obtain appropriate permissions before applying domain adaptation to sensitive speech.

\paragraph{Dual use and misuse.}
The proposed parameter-space correction could, in principle, be used to enhance ASR systems deployed in surveillance or monitoring settings. 
Our intention is to support low-resource and accessibility-oriented speech technologies, rather than enabling intrusive applications.

\section*{Acknowledgement}
This work was supported by the Ministry of Education (MOE) of Taiwan under the Taiwan Centers of Excellence in Artificial Intelligence project, through the NTU Artificial Intelligence Center of Research Excellence (NTU AI-CoRE). 
We also thank the National Center for High-performance Computing (NCHC) of the National Applied Research Laboratories (NARLabs) in Taiwan for providing computational and storage resources. 
We sincerely thank Dianna Yee, Colin Lea, and Ting-Yao Hu for their valuable feedback and insightful suggestions on this work.
\bibliography{custom}

\newpage
\appendix

\section{Model characteristic}
Table~\ref{tab:whisper_models} summarizes the key characteristics of the Whisper models used in our experiments. Parameter counts are approximate values reported by the official release.

\begin{table}[h]
\centering
\small
\begin{tabular}{lccc}
\toprule
\textbf{Model} & \textbf{Parameters (M)} & \textbf{Encoder} & \textbf{Decoder} \\
\midrule
Tiny & 39   & 4  & 4  \\
Base & 74   & 6  & 6  \\
Small & 244  & 12 & 12 \\
Medium & 769  & 24 & 24 \\
Large v2 & 1550 & 32 & 32 \\
\bottomrule
\end{tabular}
\caption{Whisper models used in our experiments. All models share the same encoder--decoder transformer architecture but differ in scale. Parameter counts are reported in millions. \textbf{Encoder} and \textbf{Decoder} are the number of transformer layers in the encoder and decoder, respectively.}
\label{tab:whisper_models}
\end{table}

\begin{table*}[t]
\centering
\small
\begin{tabular}{lcccccccccc}
\toprule
\textbf{Split} & \textbf{Igbo} & \textbf{Swahili} & \textbf{Hausa} & \textbf{Zulu} & \textbf{Twi} & \textbf{Yoruba} & \textbf{Ijaw} & \textbf{Afrikaans} & \textbf{Idoma} & \textbf{Setswana} \\
\midrule
\textbf{Train} & 8083 & 5480 & 5437 & 1306 & 1315 & 14369 & 2357 & 1911 & 1760 & 1273 \\
\textbf{Dev}   & 216  & 313  & 116  & 310  & 186  & 361   & 48   & 82   & 50   & 215  \\
\textbf{Test}  & 355  & 521  & 196  & 175  & 58   & 648   & 80   & 54   & 60   & 97   \\
\bottomrule
\end{tabular}
\caption{Number of samples per split for each accent that we used for our experiment.}
\label{tab:split_stats}
\end{table*}

\section{Data, Artifacts, and Licensing}
\paragraph{Dataset statistic} 
Table~\ref{tab:split_stats} summarizes the number of samples for each language across the train, development, and test splits.

\paragraph{Licenses and terms of use.}
All datasets and pretrained models used in this work are publicly available for research purposes under open licenses. 
We use the \textsc{AfriSpeech-200} corpus \cite{afrispeech_200}, which is distributed under a  Creative Commons Attribution NonCommercial ShareAlike v4.0 (CC BY-NC-SA 4.0) license. 
The Whisper models \cite{whisper} are released by OpenAI under the MIT license. 
Our use of these resources fully complies with their stated terms and intended use for non-commercial academic research. 
No additional data scraping or private data collection was conducted.
All artifacts associated with this work, including source code, trained correction vectors, fine-tuned model checkpoints, and documentation, will be released under the CC BY-NC-SA 4.0 license after acceptance.

\paragraph{Intended use and compatibility.}
All artifacts are used within the scope of their original research purpose, which is speech recognition and domain adaptation studies. 
We do not deploy or fine-tune any model for commercial, surveillance, or identification applications. 
Derived models and results are intended solely for academic analysis and benchmarking. 
Any derived artifacts that we release will include clear documentation of intended use and license terms to prevent misuse outside of research contexts. 

\paragraph{Anonymization and privacy protection.}
The \textsc{AfriSpeech-200} dataset contains anonymized speech recordings collected with participant consent. 
No personally identifiable information (PII) or metadata that could reveal speaker identity is used or released. 
We performed a manual spot-check and confirmed that no audio files or transcripts contain sensitive, offensive, or private information. 
All models were trained and evaluated locally on anonymized data, with no connection to external user data or APIs.

\section{Use of AI assistants}
This manuscript was refined with the assistance of large language models, which were used to improve clarity, grammar, and readability of the text. 
All conceptual development, experimental design, data analysis, and interpretation were conducted entirely by the authors. 
The AI assistants were not involved in generating research ideas or writing original scientific content.

\begin{table*}[t]
\centering
\small
\setlength{\tabcolsep}{6pt}
\renewcommand{\arraystretch}{1.25}
\begin{tabular}{p{15.4cm}}
\toprule
\textbf{Example and Description} \\
\midrule
\textbf{Example 1} \\[3pt]
\textbf{Training set Ground Truth:} Emmanuel Opuru \textcolor{green!50!black}{said} the suspects will face murder charge after investigation are complete. \\
\textbf{Teacher Pseudo-label:} the man on the opposite \textcolor{red}{side} of the suspect with face, mother, child after investigation are complete. \\
\textbf{Error Type:} Accent-induced lexical confusion—teacher model mishears ``said'' as ``side'' due to strong accent variation. \\[3pt]
\midrule
\textbf{Testing set Ground Truth:} It was a great pleasure, an audience member said later. \\
\textbf{Pretrained Tiny:} it was a great page and audience will see the later \\
\textbf{Tiny (Student):} it was a great pleasure and audience will be a \textcolor{red}{side} later \\
\textbf{Tiny (Pseudo2Real):} it was a great pleasure and audience member \textcolor{green!50!black}{said} later \\
\textbf{Error Mitigation:} Pseudo2Real effectively suppresses the accent-induced error by restoring the correct lexical item (``said''), aligning the transcription with the intended semantic meaning. \\[3pt]
\midrule
\textbf{Example 2} \\[3pt]
\textbf{Training set Ground Truth:} In figure skating, sometimes women or men skate alone, or they skate in couples. \\
\textbf{Teacher Pseudo-label:} In figure skating, sometimes women or men skate alone or they skate in couples \textcolor{red}{full stop} \\
\textbf{Error Type:} Teacher hallucination—an extra \textcolor{red}{``full stop''} token is added. \\[3pt]
\midrule
\textbf{Testing set Ground Truth:} In Nigeria, too, the May Day celebrations also happen.\\
\textbf{Pretrained Tiny:} in nigeria 2 the media celebrations also happen 1st \\
\textbf{Tiny (Student):} in nigeria 2 the medial celebrations also happen \textcolor{red}{full stop} \\
\textbf{Tiny (Pseudo2Real):} in nigeria 2 the \textcolor{green!50!black}{may day} celebrations also happen \\
\textbf{Error Mitigation:} Pseudo2Real removes the inherited hallucinated ``full stop'' and restores correct lexical content (“May Day”). \\[3pt]
\bottomrule
\end{tabular}
\caption{Examples showing how \ours corrects systematic pseudo-label errors, including accent-induced confusion (Example 1) and hallucinated tokens (Example 2).  
\textcolor{red}{Red} = error; \textcolor{green!50!black}{green} = correct token.}
\label{tab:additional_case_study}
\end{table*}

\section{Language families of the ten accents}
\label{sec:language_family}
Although all ten accents used in our experiments originate from African regions, they belong to distinct and often unrelated linguistic families, each with different phonological and prosodic characteristics. Table~\ref{tab:language_family} lists each accent together with its language family and subfamily/branch. These groups differ markedly in tonal systems, syllable structures, consonant inventories, and morphology. Therefore, while the accents are geographically African, they are not linguistically homogeneous, and adaptation across these families remains highly non-trivial.

\begin{table}[h]
\centering
\small
\setlength{\tabcolsep}{5pt}
\begin{tabular}{lll}
\toprule
\textbf{Language} & \textbf{Family} & \textbf{Subfamily / Branch} \\
\midrule
Igbo      & Niger--Congo   & Volta--Niger \\
Swahili   & Niger--Congo   & Bantu (Sabaki) \\
Hausa     & Afro-Asiatic   & Chadic \\
Zulu      & Niger--Congo   & Bantu (Nguni) \\
Twi       & Niger--Congo   & Kwa (Akan) \\
Yoruba    & Niger--Congo   & Volta--Niger \\
Ijaw      & Niger--Congo   & Ijoid \\
Afrikaans & Indo-European  & Germanic \\
Idoma     & Niger--Congo   & Volta--Niger (Idomoid) \\
Setswana  & Niger--Congo   & Bantu (Sotho--Tswana) \\
\bottomrule
\end{tabular}
\caption{Language family and subfamily/branch for each of the ten accents used in our experiments. The ten accents span three major language families (Niger--Congo, Afro-Asiatic, Indo-European) and numerous subfamilies, highlighting that the evaluation is not linguistically homogeneous despite the shared geographic origin.}
\label{tab:language_family}
\end{table}

\section{K-Means Implementation}
For the speaker clustering procedure used in the Pseudo2Real-SC variant, we employ the standard K-means algorithm from the scikit-learn library \cite{10.5555/1953048.2078195} with default configuration. The initialization method is set to \texttt{k-means++} to improve convergence speed and stability. The maximum number of iterations per run is fixed at 300, and the convergence tolerance is set to $10^{-4}$. The standard Lloyd’s algorithm is used as the clustering method. 

\section{Baseline Implementation Details}
\label{app:baseline_details}
 
\paragraph{Error Correction (EC).}
We fine-tune a T5-base model to perform post-hoc correction of ASR hypotheses. To preserve the unsupervised domain adaptation (UDA) setting, in which no target-domain ground-truth labels are available, all T5 supervision is drawn exclusively from the source domain. Concretely, we construct training pairs \textit{(hypothesis, ground-truth)} by running the Whisper teacher on source-domain speech (the five source accents of each cross-fold split) and pairing the resulting hypotheses with the corresponding human transcripts. The T5 model is thus trained to map noisy, teacher-produced hypotheses back to their reference transcripts; to the extent that the teacher makes similar mistakes on source and target speech, the learned mapping transfers to cleaning target-domain pseudo-labels at inference. Input and target sequences are truncated or padded to 256 tokens. We use a per-device batch size of 4 with 4-step gradient accumulation (effective batch size 16), a learning rate of $3\times10^{-4}$, and 500 warmup steps. The model is trained for up to 20{,}000 optimization steps with evaluation every 1{,}000 steps using cross-entropy loss over target tokens, and the checkpoint with the lowest validation loss is selected.
 
At inference, the trained T5 model is applied to the teacher-generated pseudo-labels on target-domain speech, producing a corrected pseudo-label set. The Whisper student is then fine-tuned on the corrected target-domain data \textit{(target speech, T5-corrected pseudo-label)}, following the same recipe as $\theta_{t}^{\text{pseudo}}$ but with the raw pseudo-labels replaced by their T5-corrected counterparts. EC therefore isolates the effect of text-level pseudo-label cleaning, and since T5 is trained only on source-domain supervision and applied to target-domain pseudo-labels without access to any target ground truth, this baseline remains within the UDA setting.
 
\paragraph{Exponential Moving Average (EMA).}
EMA is applied on top of the $\theta_{t}^{\text{pseudo}}$ baseline, which fine-tunes the Whisper student on target-domain pseudo-labels. Let $\theta^{(k)}$ denote the student's parameters at training step $k$ and $\hat{\theta}^{(k)}$ the EMA parameters, initialized as $\hat{\theta}^{(0)} = \theta^{(0)}$. After each optimizer update, the EMA weights are updated as
\[ \hat{\theta}^{(k)} = \beta\,\hat{\theta}^{(k-1)} + (1-\beta)\,\theta^{(k)}, \]
with decay factor $\beta = 0.999$, which yielded the best average validation performance among the values we tried. The EMA parameters are maintained separately from the training weights and do not participate in gradient computation. During evaluation and checkpoint saving, the training weights are temporarily swapped with the EMA weights and restored immediately afterward to continue optimization. All other training details are identical to the $\theta_{t}^{\text{pseudo}}$ baseline; in particular, no target-domain ground-truth labels are used, preserving the UDA setting.

\section{Additional case study}
To further illustrate how \ours mitigates systematic pseudo-label errors, we provide more qualitative examples in Table~\ref{tab:additional_case_study}. These cases highlight two common types of errors in teacher-generated pseudo-labels and show how \ours effectively addresses them.

\textbf{Example 1} illustrates an accent-induced lexical confusion. Here, the teacher model mishears “said” as “side” due to strong accent variation in the training data, producing semantically inconsistent pseudo-labels. This error propagates to the student model, which reproduces the mistaken “side” in the testing example, leading to incorrect transcription. \ours successfully corrects this accent-induced error by restoring the intended token “said,” aligning the transcription with the ground truth.

\textbf{Example 2} shows that the teacher model introduces an extraneous token (“full stop”) which does not exist in the ground-truth transcription. Such hallucinated tokens often arise from overconfident language modeling behavior and can propagate into the student model trained on these pseudo-labels. When the same error type appears in the testing example, the student model reproduces this pattern, again appending a spurious “full stop” at the end.

By contrast, \ours completely removes the hallucinated “full stop” pattern, demonstrating that the erroneous token sequence no longer appears in the output. Moreover, it restores the correct lexical content (“May Day”) that aligns with the ground-truth transcription. This indicates that the parameter-space correction in \ours not only suppresses inherited hallucinations but also reinforces meaningful acoustic-text alignment, leading to more faithful and semantically accurate transcriptions than direct pseudo-label fine-tuning.



\end{document}